\documentclass[10pt,showpacs,showkeys,preprint]{revtex4-1}
\usepackage{bm}
\usepackage{amssymb}
\usepackage{amsmath}
\usepackage{graphicx}
\usepackage{multirow}
\usepackage{MnSymbol}
\usepackage{wasysym}
\usepackage{stmaryrd}
\usepackage[outline]{contour}
\usepackage[dvipsnames]{xcolor}
\usepackage{float}
\usepackage[mathlines]{lineno}

\definecolor{olive}{RGB}{182,187,37}

\begin{document}

\title{Atomic ionization by multicharged ions interpreted in terms of poles
in the charge complex space}
\author{J. E. Miraglia}
\date{\today }

\begin{abstract}
In this work we study the single ionization of hydrogen and helium by the
impact of a punctual Coulomb projectile. To interpretate the cross section
we introduce a series of Pad\'{e} approximant. The nodes of the denominator
of the Pad\'{e} approximant give rise to poles in the charge complex plane.
These poles move with the projectile velocity following a certain pattern.
We find the positions of the poles of the Continuum Distorted Wave theory
and an ensamble of experimental results. It was found that the ionization
cross section may have oscillations in terms of the incident charge for a
given impact velocity.
\end{abstract}

\maketitle

\affiliation{Instituto de Astronom\'{\i}a y F\'{\i}sica del Espacio. Consejo Nacional de
Investigaciones Cient\'{\i}ficas y T\'{e}cnicas. 
 \\
Casilla de Correo 67, Sucursal 28, {C1428EGA} Buenos Aires, Argentina.}

In Ref.\cite{Miraglia2020a}, we have studied the \ single ionization cross
section $\sigma (Z,v)\ $by impact of a punctual Coulomb charge $Z$ on
hydrogen and helium as a function of the impinging velocity $v$.\ In that
article, we proposed that the cross section can be separated in the
asymptotic limit times a Pad\'{e}[8,8] written in terms of correlated eight
poles in the \textit{velocity complex plane}: four in the upper complex
plane and their conjugate ones in the lower plane. Making use of a \
numerical data set (NDS) of 443 numerical calculations of the Continuum
distorted wave eikonal initial state (CDW for short), and contrasting them
with\ an experimental data set (EDS) of 328 values, we estimated the
validity range of the CDW theory, the position of the poles predicted by the
CDW \ and also the poles replicating the experiments. In this article, we
tackle the problem the other way around, i.e. we find an explicit form of
the ionization cross section $\sigma (Z,v)$ as a function of $Z$ for a given
impact velocity $v$ by finding the positions of the poles in the\textit{\
charge complex\ plane}.

\section{Theory}

Following the same scheme as in Ref.\cite{Miraglia2020a}, we here resort to
a series of Pad\'{e}[$2j-2,2j$] to modulate the Born approximation $\sigma
^{B}(Z,v)$, with the following structure 
\begin{equation}
\sigma ^{\lbrack J]}(Z,v)=\sigma
^{B}(Z,v)\dsum\nolimits_{j=1}^{J}c_{j}P_{2j-2,2j}(Z),  \label{100}
\end{equation}%
where the Pad\'{e} approximants are defined in terms of poles in the complex
plane of the charge at%
\begin{equation}
Z_{\pm j}=z_{jr}\pm iz_{ji},  \label{110}
\end{equation}%
where $z_{jr}$ and $\pm z_{ji}$ are the real and imaginary components of the
pole $Z_{\pm j},$ and $c_{j}$ is its corresponding strength.$\ $The Pad\'{e}
terms of Eq.(\ref{100}) are defined as follows%
\begin{eqnarray}
P_{0,2}(Z) &=&\frac{\left\vert Z_{1}\right\vert ^{2}}{\left\vert
Z-Z_{1}\right\vert ^{2}},  \label{120} \\
P_{2,4}(Z) &=&\frac{Z^{2}\ \left\vert Z_{2}\right\vert ^{2}}{\left\vert
Z-Z_{1}\right\vert ^{2}\left\vert Z-Z_{2}\right\vert ^{2}},  \label{130} \\
P_{4,6}(Z) &=&\frac{Z^{4}\ \left\vert Z_{3}\right\vert ^{2}}{\left\vert
Z-Z_{1}\right\vert ^{2}\left\vert Z-Z_{2}\right\vert ^{2}\left\vert
Z-Z_{3}\right\vert ^{2}}  \label{140}
\end{eqnarray}%
and so on. Similarly to Ref.\cite{Miraglia2020a}, we cast all the
information in the position of the poles in the complex plane of the
projectile charge, i.e. on $z_{jr}$ and $z_{ji},$ and on the strength $%
c_{j}, $ but now these magnitudes depend on the\ impact velocity $v.$ As
before, for each pole $Z_{j}$ in the upper plane there is another one
conjugated in the lower one$\ Z_{-j}=Z_{+j}^{\ast }$. The strength of the
first pole must be forced to be unity, i.e. $c_{1}=1,$ to satisfy 
\begin{equation}
\underset{Z\rightarrow 0}{\lim }\ \sigma ^{\lbrack J]}(Z,v)=\sigma
^{B}(Z,v)=Z^{2}\sigma ^{B}(1,v),  \label{150}
\end{equation}%
which is the correct perturbative limit. And the only one that we know as
certain, because there is no clue about the behavior for large $Z.\ $At
large perturbation $\sigma ^{\lbrack J]}(Z,v)$ saturates order to order
differently, that is%
\begin{equation}
\sigma ^{\lbrack J]}(Z,v)\underset{v\rightarrow \infty }{\rightarrow }\sigma
^{B}(1,v)\left\{ 
\begin{array}{ll}
\left\vert Z_{1}\right\vert ^{2},\ \  & \text{ for }J=1 \\ 
\left\vert Z_{1}\right\vert ^{2}+c_{1}\left\vert Z_{2}\right\vert ^{2},\ \ 
& \text{ for }J=2 \\ 
\left\vert Z_{1}\right\vert ^{2}+c_{1}\left\vert Z_{2}\right\vert
^{2}+c_{3}\left\vert Z_{3}\right\vert ^{2},\ \  & \text{ for }J=3%
\end{array}%
\right.  \label{160}
\end{equation}%
and so on. One would expect the poles to be\ ordered in terms of their of
modulus, i.e\ $|Z_{a}|^{2}<|Z_{b}|^{2}<|Z_{c}|^{2}$ and they come into the
picture in accordance with the increasing value of$\ Z$.

The zero order of this succession can be defined as $\sigma ^{\lbrack
0]}(Z,v)=\sigma ^{B}(Z,v)$ stressing the importance of the \textit{full}
first Born approximation to provide the proper limit. By \textit{full} we
mean that its calculation covers\ properly the whole range of velocity; not
just the asymptotic limit as in Ref.\cite{Miraglia2020a}.

\subsection{The validity regime}

Within the perturbative regime, defined now as $Z<|Z_{a}|,$ Born
approximation $\sigma ^{B}(Z,v)$ holds. \ As $Z$ increases, larger poles
start to play decisive roles. Before facing the task of finding the
positions of the poles, we will focus on the range of validity of the CDW.
By using the pool of CDW calculations (see numerical data set, Figure 1c and
2c of Ref.[1]) we defined $v_{\max }\ $where the cross section is maximum: $%
\sigma _{\max }=\sigma (v_{\max })$ to give%
\begin{equation}
v_{\max }^{2}\approx \left( v_{\max }^{B\ \ }\right) ^{2}+c_{1}Z,
\label{200}
\end{equation}%
where $v_{\max }^{B\ \ }$ =$1.\ \ (1.25)$ is the \ velocity where the Born
approximation is maximum for hydrogen (helium) and the remaining coefficient
was fitted to be\ \ $c_{1}=1\ (1.59)$\ for hydrogen (helium). This relation
was fundamental, since it allowed us to introduce a criterion to define the
validity of the CDW-theory, as: $v>v_{\max }$ ,\ and most of the conclusions
of Ref. \cite{Miraglia2020a} were digested in terms of the ratio $v/v_{\max
} $. But here we should invert this relation since we are dealing with $Z$
as a variable. The criterion should change to%
\begin{equation}
Z<Z_{\max }=\frac{v^{2}\ -\left( v_{\max }^{B\ \ }\right) ^{2}}{c_{1}},
\label{210}
\end{equation}%
and to express any finding in terms of the ratio $Z/Z_{\max }$. With this
definition we can see differently Figures \ 1a and 2b\ of Ref. \cite%
{Miraglia2020a} by replotting the magnitude%
\begin{equation}
e^{CDW-\exp }=\frac{\sigma ^{CDW}(Z,v)-\sigma ^{\exp }(Z,v)}{\sigma ^{\exp
}(Z,v)}\times 100,  \label{215}
\end{equation}%
but now\ as a function of $Z$/$Z_{\max }$, as it is shown in Figures \ 1a
and 1b of the present article for hydrogen and helium, respectively. If the
validity of the CDW in Ref. \cite{Miraglia2020a} was expressed in the domain 
$v/v_{\max }>1,$ here it is translated as: $Z/Z_{\max }<1.$ 
\subsection{The poles of the CDW theory}

Next, we proceed to find the values of the first two poles, $Z_{1}$ and $%
Z_{2}$ governing the $s^{[1]}$ and $s^{[2]}\ $\ using the pool of CDW
numerical calculation (see NDS of Ref.[1]).\ In Figure 2 we plot the\
normalized magnitude%
\begin{equation}
s^{[J]}[Z,v]=\frac{\sigma ^{\lbrack J]}(Z,v)}{\sigma ^{B}(Z,v)},  \label{220}
\end{equation}%
$s^{[1]}$ in dashed blue$,$ and $s^{[2]}$ in \ solid red as a function of
the impinging charge $Z$ along with the numerical CDW values shown with
empty green circles for helium targets. Four impact energies were
considered: 1000, 500, 250 an 100\ kev/amu. Note that in this graphic $%
s^{[J]}=1$ represents the Born approximation, so any departure from unity
accounts for the distorted wave contribution. It is important to note that $%
Z_{1}$ obtained for $s^{[1]}$ does not coincide precisely with the one of $%
s^{[2]},$ because each order introduces a new pole but corrects accordingly
the position of the previous ones. We design a fitting procedure so the
poles of $s^{[J]}\ $are the seeds for the new generation of poles for $%
s^{[J+1]}$, in this way the new ones are derived with the knowledge of their
ancestors.

From Figure 2, we conclude that $s^{[1]}$\ reproduces the CDW in the reduced
range $Z<<Z_{\max },$ while$\ s^{[2]}$ reproduces the CDW numerical value in
almost the whole $Z$ regime, including $Z\sim Z_{\max }.$ To express this
convergence in numbers, we display in Figure 1c-f the relative errors%
\begin{equation}
e^{[J]-CDW}=\frac{\sigma ^{\lbrack J]}(Z,v)-\sigma ^{CDW}(Z,v)}{\sigma
^{CDW}(Z,v)}\times 100  \label{230}
\end{equation}%
as a function of $Z/Z_{\max }$ for hydrogen and helium and for $J$=1 and $J$%
=2. The two-pole order $\sigma ^{\lbrack 2]}$ represents the numerical CDW
values within very few percents which means that we find a successful \
approach. There is no point in proceeding with a third pole because we will
be dealing with a region far beyond its validity

It is quite interesting to display the position of the pole $z_{1r}$ and $%
z_{1i}$, \ shaping $\sigma ^{\lbrack 1]}$ in terms of $v$ as shown in Figure
3 for hydrogen and helium. They follow a pattern, which can be fitted with a
simple expression depending on just three parameters; for example the
fitting functions displayed in the figure read%
\begin{equation}
z_{1r}=\frac{a+bv}{1+c(v-1)^{2}},\text{ and \ \ }z_{1i}=d+ev+fv^{2}
\label{225}
\end{equation}%
Also in the Figure we display the value of \ $Z_{\max }$ to indicate the
range of validity: In all cases $|Z_{1}|<|Z_{\max }|$. \ . 

\subsection{The poles of the experiments}

Following the same structure as Ref. \cite{Miraglia2020a}, we would like to
find the positions of the poles replicating the experimental data. \ But, we
find that it is rather impossible to put together a set of data for a given $%
v$ in items of $Z$ sufficiently large to be reasonably fitted. We find just
one case: 100 keV/amu impact on helium where we can put together the results
of antiprotons Refs. \cite{Andersen1990,Hvelplund1994}, $H^{+},\ He^{++}$
and Li$^{3+}$ by the Belfast group \cite{Shah1985}, and the results of Datz
and collaborators \cite{Datz1990} with $Z\ $\ ranging from 5 to 16,
totalizing \ just twelve points, as shown in Figure 4. Born approximation at
this impact energy is $\sigma ^{B}=$3.34$Z^{2}$ in atomic units and $Z_{\max
}=1.1.$ So, except protons and antiprotons impact, the impinging charges are
quite outside the validity range, and indeed the CDW theory as shown in the
figure collapses for larger charges. \ With this rather scarce experimental
data, we obtain $\sigma ^{\lbrack 1]},\ \sigma ^{\lbrack 2]},$ and $\sigma
^{\lbrack 3]}$ as plotted in the figure which describes quite well the
experimental data available. It seems that each pole relates to one
oscillation. As we would include more poles, it is possible that the cross
section would continue oscillating for larger charges. It would give rise to
a very interesting behaviour which should be related to the influence of the
\ channels of capture.

The idea that the ionization cross\ section presents oscillations for $%
Z>Z_{\max }$ is a very interesting one, that requires some experiments to be
confirmed. If this were the case it is possible that the poles follow some
patterns related to the capture mechanism. We should mention here some
resemblances, such as the oscillations of the reflectance coefficients in a
collision with a squared well, \ or the oscillation of \ the excitation
probability of the collision of two one-dimensional squared wells, when we
plotted in terms of the depth of the moving well \cite{Rodriguez1991}.

\section{Bibliography.}

{}

\begin{figure*}[t]
\centering
\includegraphics[width=0.90\textwidth]{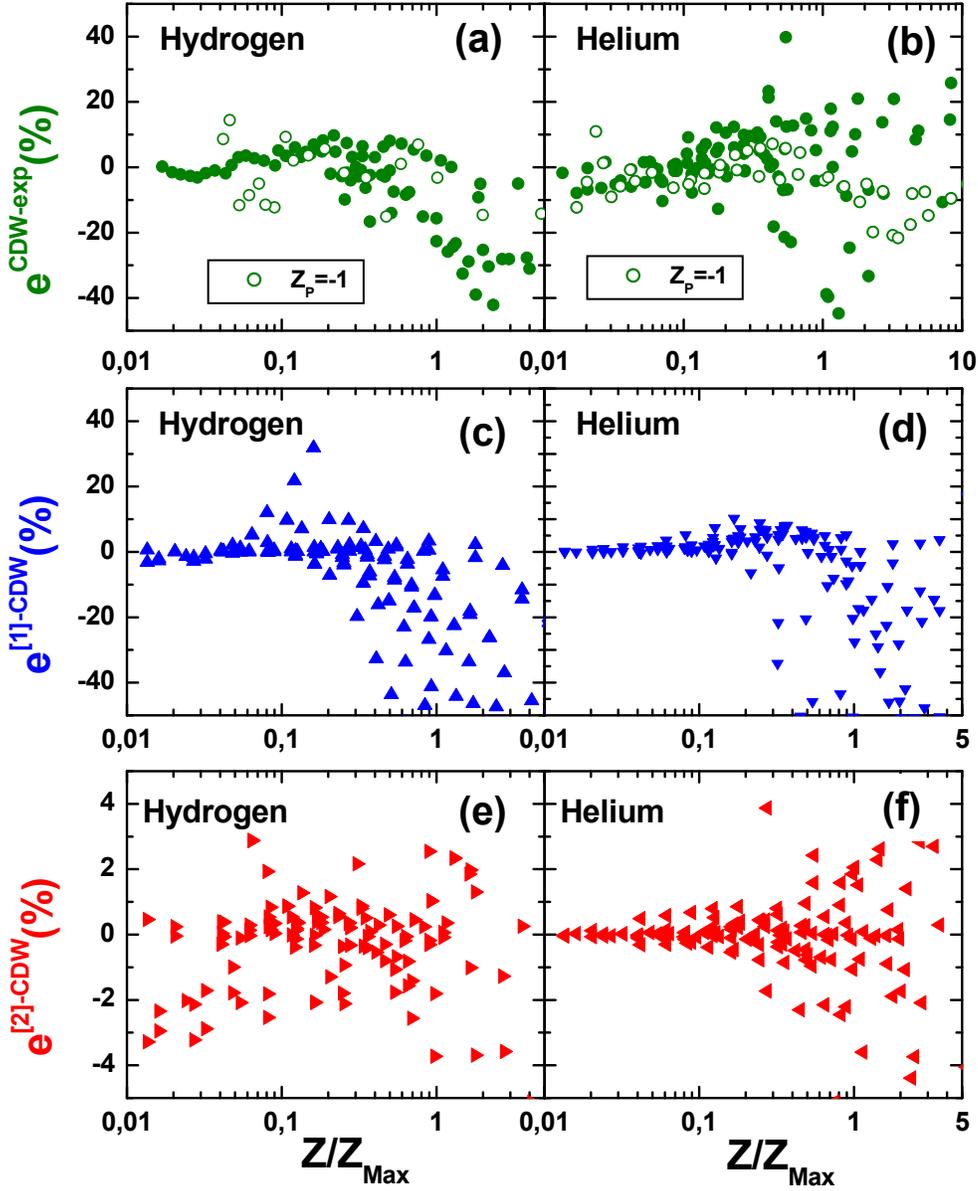}
\caption{(Color online) Hydrogen target. (a) Relative error of the CDW
versus the experiments as defined in Eq.(10) in terms of $Z/Z_{max}$. (c)
Relative error of $\protect\sigma ^{[1]}$ with respect to the CDW as defined
in Eq.(13) (e) Similar to (c) for $\protect\sigma ^{[2]}$. (b), (d) and (e)
similar to (a), (c) and (f) for helium target.}
\end{figure*}

\begin{figure*}[t]
\centering
\includegraphics[width=0.90\textwidth]{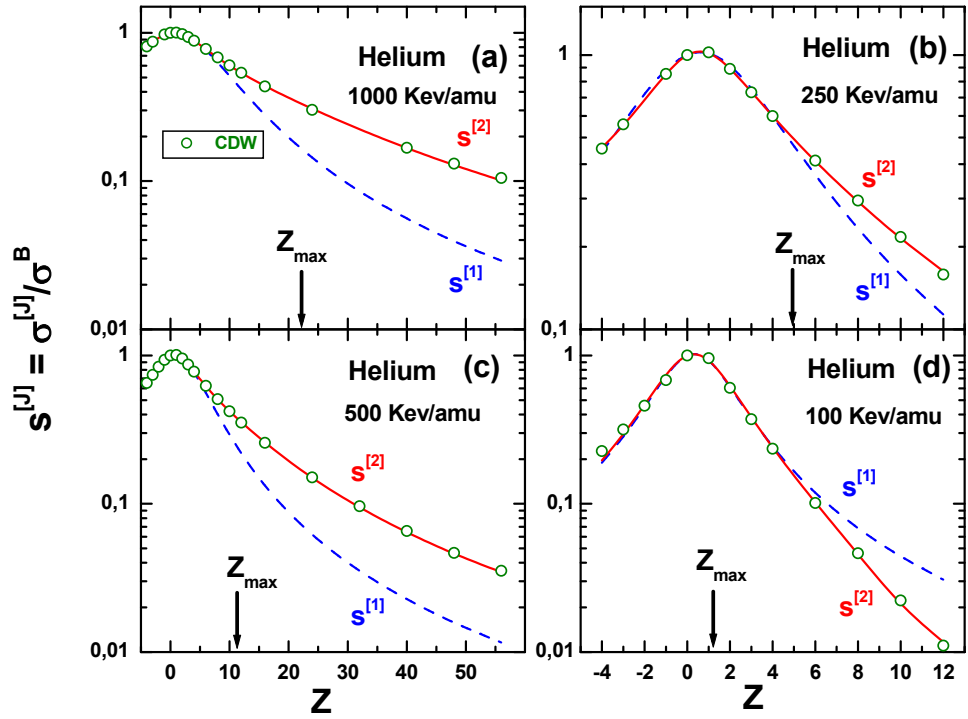}
\caption{(Color online) Values of $s^{[1]}$ and $s^{[2]}$, as defined in
Eq.(11) as a function of the Coulomb incident charge on helium for four
impinging energies, as indicated.}
\end{figure*}

\begin{figure*}[t]
\centering
\includegraphics[width=0.90\textwidth]{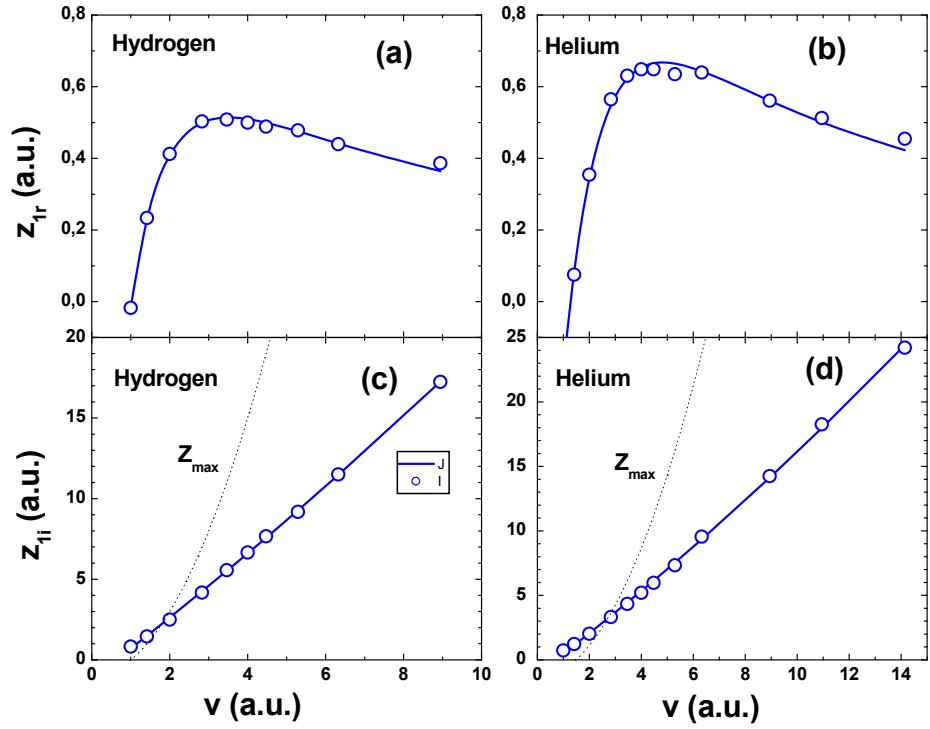}
\caption{(Color online) Positions of the real and imaginary parts of the
pole $Z_{1}$ of $\protect\sigma ^{^{\lbrack }[1]}$, as a function of the
impact velocity for hydrogen and Helium as indicated. The symbols are the
numerical values and the solid lines are the fitting given by Eqs(12).}
\end{figure*}

\begin{figure*}[t]
\centering
\includegraphics[width=0.90\textwidth]{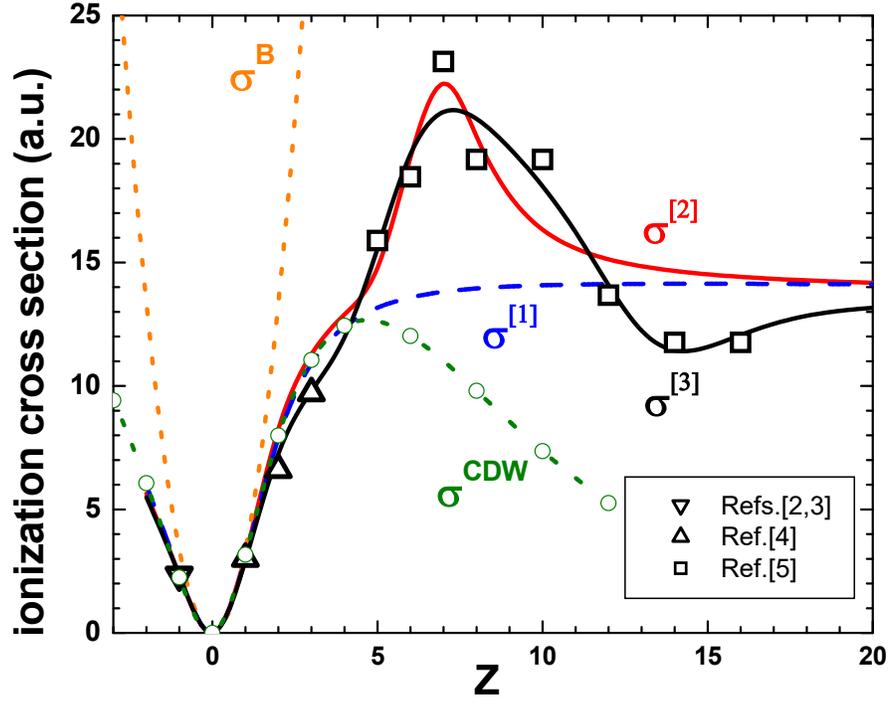}
\caption{(Color online) Ionization cross section of helium as a function of
the incident Coulomb charge for 100kev/amu impact energy. The black empty
symbols are the experiments, and first Born approximation, CDW, $\protect%
\sigma^{[1]}$, $\protect\sigma^{[2]}$, and $\protect\sigma^{[3]}$ as noted.}
\end{figure*}

\end{document}